\documentclass[aps,preprint,groupedaddress]{revtex4-1}

\usepackage{amsmath} 
\usepackage{amssymb}	
\usepackage{graphicx} 
\newcommand{\wt}[1]{\ensuremath{\tilde{#1}}}
\renewcommand{\v}[1]{\ensuremath{\mathbf{\overline{#1}}}} 
\newcommand{\pd}[2]{\frac{\partial #1}{\partial #2}} 
\newcommand{\pdd}[2]{\frac{\partial^2 #1}{\partial #2^2}} 

\begin{document}

\title{ A New Approach to Cascaded Stimulated Brillouin Scattering}

\author{Mark Dong}
\email{markdong@umich.edu}

\author{Herbert Graves Winful}
 \email{arrays@umich.edu}
\affiliation{Department of Electrical Engineering and Computer Science, University of Michigan, \\1301 Beal Avenue, Ann Arbor, 48109-2122}

\date{\today}

\begin{abstract}
We present a novel approach to cascaded stimulated Brillouin scattering and frequency comb generation in which the multitude of interacting pump, Stokes, and anti-Stokes optical fields are described by a single forward wave and a single backward wave at a single carrier frequency.  The envelopes of these two waves are modulated through coupling to a single acoustic oscillation and through four-wave mixing.  Starting from a single pump field, we observe the emergence of a comb of frequencies as the intensity is increased.  The set of three differential equations derived here are sufficient to describe the generation of any number of Brillouin sidebands in oscillator systems that would have required hundreds of coupled equations in the standard approach.  We test the new approach on some published experiments and find excellent agreement with the results. 

\end{abstract}


\maketitle

\section{Introduction}

        Stimulated Brillouin scattering (SBS) transfers energy from a strong optical pump wave to a lower frequency Stokes wave through an interaction with a hypersonic acoustic wave when the pump power exceeds a certain threshold \cite{Chiao1964, Boyd2008, Agrawal2007, Kob2010}.  In principle, when the Stokes wave also exceeds the threshold for SBS it can generate a higher order Stokes wave.  In practice, higher order Stokes scattering is rarely seen in the absence of feedback because the threshold for even the second order is two orders of magnitude higher than that for the first\cite{Russel2002}.  This is due to the fact that the effective length of the medium occupied by the first order Stokes is extremely small and localized near the entrance.  The presence of reflective feedback or a resonant cavity dramatically reduces the threshold for higher order scattering thus making it possible to observe a cascade of higher order Stokes lines\cite{Hill1976}.  Anti-Stokes orders can also be generated through four-wave mixing once there are at least two co-propagating pump and Stokes waves\cite{Labudde1980}.  The result is a large number of equally-spaced lines that are coherently phased: a Brillouin frequency comb \cite{Braje2009, Li2012, Buttner2014-1, Buttner2014-2, Lin2014, Loranger2012}.  These frequency combs have attracted a great deal of interest because of their potential applications in areas such as microwave photonics, sensing, spectroscopy, and wavelength-division multiplexing \cite{Kippenberg2011}. 
        
        In the theoretical analysis of cascaded SBS, all previous studies have treated the Stokes shifted waves as separately propagating waves, each with its own frequency and wave vector coupled to their co-propagating acoustic waves \cite{Randoux1995, Ogusu2002-1, Ogusu2002-2, Ogusu2003}.  In a comb with many frequencies this approach can lead to an enormous number of coupled equations that need to be solved simultaneously.  For example, with just 3 Stokes orders, Ogusu had to solve 14 coupled partial differential equations for the forward and backward propagating pump, Stokes, and acoustic waves \cite{Ogusu2002-1}.  Indeed there are experimental reports of Brillouin frequency combs with as many as 800 lines\cite{Min2001}.  A detailed numerical study of such a comb would require solving more than 3200 partial differential equations.  In the presence of four-wave mixing, anti-Stokes components will also be generated and thus the number of coupled equations will blossom rapidly. 
	
        These previous studies miss one essential feature of SBS, the fact that the Brillouin frequency shift, of order 10 GHz in silica, is four orders of magnitude smaller than the carrier frequency of the optical waves involved.  The generated Stokes process can be regarded then as an acoustic modulation on top of the carrier waves.  With that recognition the entire process of cascaded SBS can be seen as the interaction between two counter-propagating light waves modulated by the acoustic vibrations and the four-wave mixing process.  
	
        In this paper we introduce a new way to study cascaded SBS processes that lead to the formation of frequency combs.  Instead of solving potentially thousands of coupled wave equations for the interacting optical and acoustic modes, we reduce the problem to the solution of just three equations in three variables: a forward optical wave, a backward optical wave, and a single acoustic wave described by a second order differential equation in time.  With this approach one does not manually introduce the higher order Stokes waves.  They emerge naturally with their appropriate frequency shifts as the pump intensity is increased.  With this new approach we simulate some published experiments and obtain excellent agreement with those results.

\section{Theory}

In the traditional approach the electric field of the many light waves interacting in a Brillouin-active medium is taken as the superposition of waves of the form 

\begin{align*}
E_j(\v{r},t) = F_j(x,y) A_j(z,t) e^{i(k_j z-\omega_j t)}
\end{align*}

\noindent in which each frequency component $\omega_j$ has its own wave vector $k_j$ and envelope $A_j(z,t)$.  Here $\omega_j=\omega_0\mp j\Omega_B$, $\omega_0$ is the pump frequency and $\Omega_B$ is the acoustic frequency.  In our approach we extract a single carrier frequency $\omega_0$ and its associated wave vector $\pm k_0$ and lump all the frequency shifted components into a single envelope $A_+(z,t)$ for the forward propagating field and $A_-(z,t)$ for the backward propagating field.  For this to be valid, the wave vector $k_0$ must not change significantly over the bandwidth of the envelope.  This condition is indeed satisfied in SBS since the frequency shift ($\Omega_B/2\pi \approx$ 10 GHz in silica) is typically several orders of magnitude smaller than the carrier frequency ($\omega_0/2\pi \approx$ 100 THz).  We thus write the fields as 

\begin{subequations}
\begin{align}
E_{+}(\v{r},t) &= F_p(x,y) A_{+}(z,t) e^{i(k_0z-\omega_0t)} + c.c.\\
E_{-}(\v{r},t) &= F_p(x,y)  A_{-}(z,t)e^{-i(k_0 z+\omega_0)}+c.c.\\
 \rho(\v{r},t) &= F_A(x,y) Q(z,t) e^{iq_0z}+c.c.
\end{align}
\label{efield}
\end{subequations}

\noindent where $F_p$  and $F_A$  are the transverse profiles of the optical and acoustic fields, in the context of an optical fiber. We now have only two optical field variables, a forward propagating and a backward propagating wave expanded around the carrier frequency $\omega_0$. The acoustic wave variable $\rho(\v{r},t)$ represents the material density fluctuation from its average value $rho_0$. In contrast to the optical fields, we extract only the propagation constant $q_0$ and retain the rapid temporal oscillations in the acoustic envelope variable $Q(z,t)$. These variables are then used in the wave equations for the material density fluctuation

\begin{align}
\pdd{\rho}{t}-\Gamma_A \nabla^2\left(\pd{\rho}{t}\right)-v_A^2\nabla^2\rho = -\epsilon_0 \gamma_e \nabla^2(\v{E}\cdot\v{E})
\end{align}

\noindent and the optical field

\begin{align}
\nabla^2 \v{E} - \frac{\epsilon}{c^2}\pdd{\v{E}}{t} = \mu_0 \pdd{\v{P}_{NL}}{t}
\end{align}

\noindent Here, $\Gamma_A$ is a damping constant, $v_A$ is the acoustic velocity, $\gamma_e$ is the electrostriction constant. The permittivity $\epsilon = n^2 - i\alpha n/c$ includes the refractive index $n$ and the loss coefficient $\alpha$. The driving term in the optical wave equation is the nonlinear polarization 

\begin{align}
\v{P}_{NL} = \epsilon_0 \left( \frac{\gamma_e}{\rho_0} \rho \v{E} + \chi^{(3)} \v{E}\v{E}\v{E}\right)
\end{align}

\noindent where $\chi^{(3)}$ is the third order nonlinear susceptibility. The derivation follows the standard approach detailed in Refs \cite{Chiao1964, Boyd2008, Agrawal2007, Kob2010}, the only difference being that we do not identify a Stokes wave with a down-shifted frequency but consider two counterpropagating waves with the same carrier frequency $\omega_0$.  The propagation constants of the optical and acoustic waves satisfy $q_0 = 2k_0$. Using the slowly varying envelope approximation for the optical fields and keeping only the phase-matched terms, we obtain the following equations after appropriate rescaling: 
 
 \begin{subequations}
\label{main}
\begin{align}
\pd{}{z}A_{+} + &\frac{1}{v_g} \pd{}{t}A_{+} = \frac{g_B}{2A_{eff}} Q A_{-}-\frac{\alpha}{2}A_++i\gamma(|A_{+}|^2 + 2|A_{-}|^2)A_{+}\\
-\pd{}{z}A_{-} + &\frac{1}{v_g} \pd{}{t}A_{-} = -\frac{g_B}{2A_{eff}} Q^* A_{+}-\frac{\alpha}{2}A_-+i\gamma(|A_-|^2+ 2|A_+|^2)A_-\\
&\left[ \pdd{}{t} + \frac{1}{\tau_B}\pd{}{t} + \Omega_B^2\right] Q = \left(\frac{4i\Omega_B}{\tau_B}\right)A_+A_-^*+ \wt{f}
\end{align}
\end{subequations}

\noindent Here $g_B$ (m/W) is the Brillouin gain, $v_g$ is the group velocity, $A_{eff}$ is the effective area, $\gamma$ (1/Wm) is the third order nonlinear coefficient, $\alpha$ is the linear loss, $\tau_B = 1/\Gamma_B$ is the Brillouin lifetime, $\Gamma_B = q_0^2 \Gamma_A$, and $\wt{f}$ is a Langevin noise source \cite{Kob2010, Boyd1990}.  In writing down the equation for the acoustic wave we have made the usual assumption that the phonons are heavily damped and do not propagate far, hence the spatial derivatives can be dropped.  We however retain the second derivative in time and hence the acoustic wave equation corresponds to a driven, damped harmonic oscillator with resonances at $\pm \Omega_B$. The positive and negative frequencies correspond to forward and backward propagating acoustic waves. In the optical equations, the nonlinear terms that have phase-matching around frequency $\omega_0$ has reduced to only three components, two of them degenerate, to form what is typically the self-phase and cross-phase modulation terms. However, and the beauty of these equations is now apparent, these two terms contain not only phase modulations but also every possible four-wave mixing combination in the vicinity of $\omega_0$, whose gains are automatically determined by the built-in phase-matching from each envelope's frequency components. The Brillouin gain linewidth and anti-Stokes scattering also emerge from the above equations.  These equations differ from the ones commonly used in that they include only a single envelope equation for the forward wave and a single envelope equation for the backward wave.  The generation of Stokes and anti-Stokes orders will appear as modulations of the envelopes.  In addition the acoustic wave is described by a single second order differential equation.  Unlike other approaches where a second-order differential equation is used for the acoustic wave, we do not extract the acoustic frequency and hence the variable $Q$ does oscillate at a frequency of several gigahertz.

\section{Numerical Results and Discussion}

\begin{figure}[h]
\begin{center}
\includegraphics[scale=0.7]{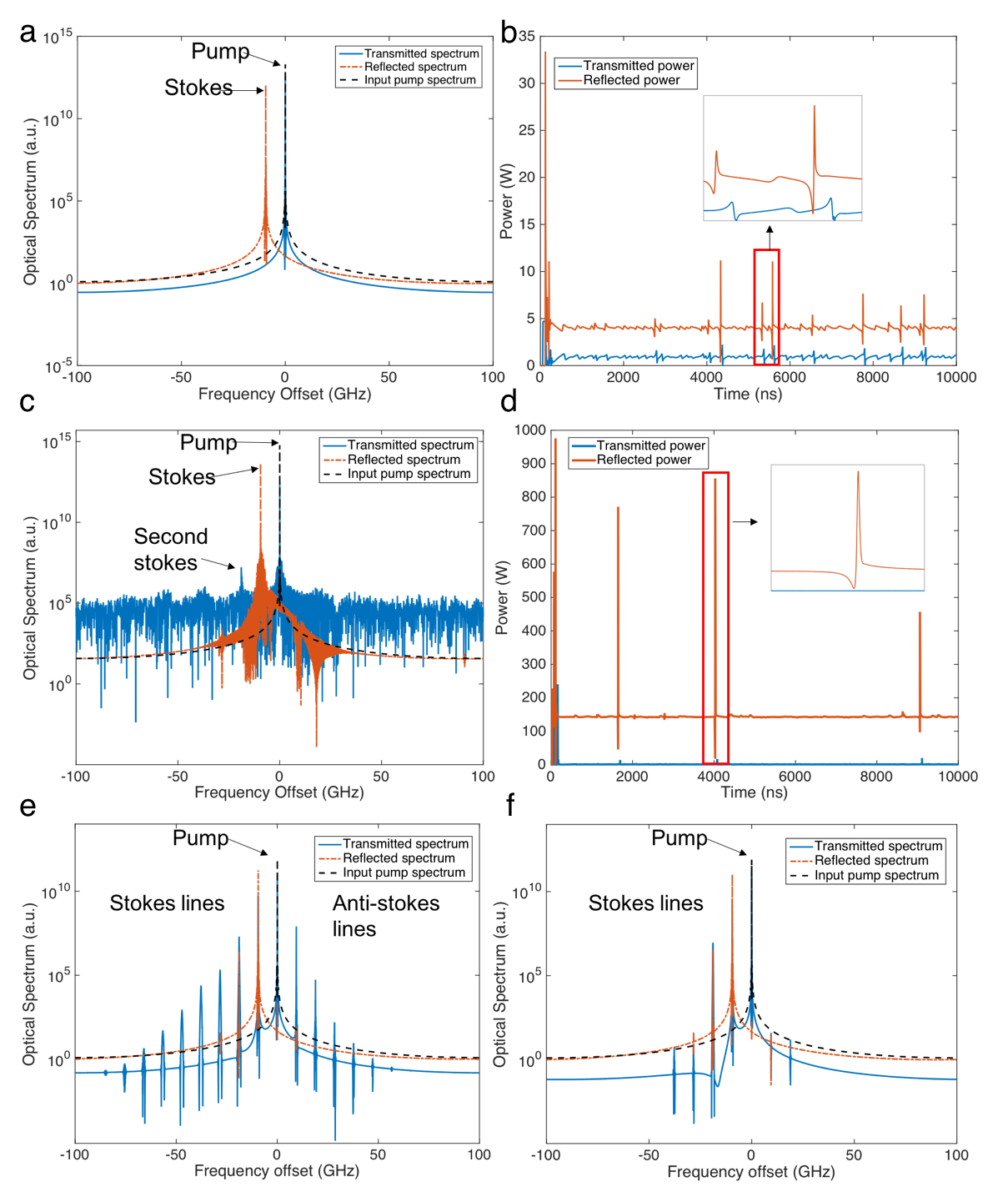}
\end{center}
\caption{(a), (b) Spectra and temporal output for a 5W cw pump in a 10-m silica fiber without facet reflections. (c), (d) Spectra and temporal output for a 150 cw pump in a 10-m silica fiber without facet reflections. Note the emergence of a second Stokes line. (e) Transmitted and reflected spectra from a fiber with a Fresnel reflecting input facet and a 5 W cw pump.  Many Stokes and anti-Stokes lines are now visible due to four-wave mixing (f) Reflected and transmitted spectra for fiber with Fresnel reflections but without four-wave mixing.  Only pump, first, and second Stokes lines are seen.}
\label{silica_fig}
\end{figure}

  Equations \ref{main} are what we use to model cascaded SBS and comb generation.  We solve them numerically via a multi-step method: integration along the characteristics for the optical equations and a fifth-order predictor-corrector algorithm (Adams-Bashforth-Moulton formulas) for the acoustic equation. Using the parameters listed in Table \ref{table_param} we first simulate a 10-meter highly non-linear silica fiber without any internal reflection from the fiber facets.  For such a fiber the threshold for first order SBS is about 1.3 W whilst the threshold for the second order Stokes is about 130 W \cite{Russel2002}.  Figure \ref{silica_fig} shows the spectra of transmitted and backscattered waves at different power levels for a continuous wave input.  Even though the input pump is single frequency, for an input of 5 W we see a strong Stokes shifted output in the reflected spectrum.  The frequency shift of 9.44 GHz is established by the acoustic oscillation of Eq. \ref{main}c.  The temporal output exhibits random bi-polar spikes due to the noise excitation.  When the pump power is raised to 150 W a weak second Stokes output is seen in the transmitted spectrum.  This second Stokes line is excited by the backward traveling first Stokes and travels in the same direction as the pump.  We stress that this second Stokes line was not inserted as an additional wave equation but emerged naturally as a modulation on top of the forward traveling input wave.  In the time domain the random spikes are much stronger and have the appearance of the extreme events known as rogue waves \cite{Hammami2008}.  We now introduce feedback for the first order Stokes wave by making the front facet 20\% reflective.  The input power is also reduced from the 150 W used in Fig. \ref{silica_fig}c, \ref{silica_fig}d to a mere 5 W.  With feedback the threshold for the higher-order scattering is greatly reduced\cite{Russel2002}. The reflected first order Stokes is now able to excite a strong second order Stokes traveling in the backward direction.  Four-wave mixing between co-propagating waves in the forward direction then gives rise to multiple anti-Stokes and Stokes orders.  It is remarkable that all these lines emerge naturally from the solution of just the three equations.  To confirm the role of four-wave-mixing we turn off the third-order nonlinearity in Eqs. \ref{main} by setting $\gamma = 0$.  Fig \ref{silica_fig}f shows that the only lines remaining in the spectrum are now the pump, first, and second order Stokes, along with a few weak lines 10-orders of magnitude lower that probably arise from photon-phonon mixing processes with no gain.  This also confirms that the terms which have the form of self-phase and cross-phase modulation are indeed responsible for four-wave mixing.    

\begin{figure}[h]
\begin{center}
\includegraphics[scale=0.7]{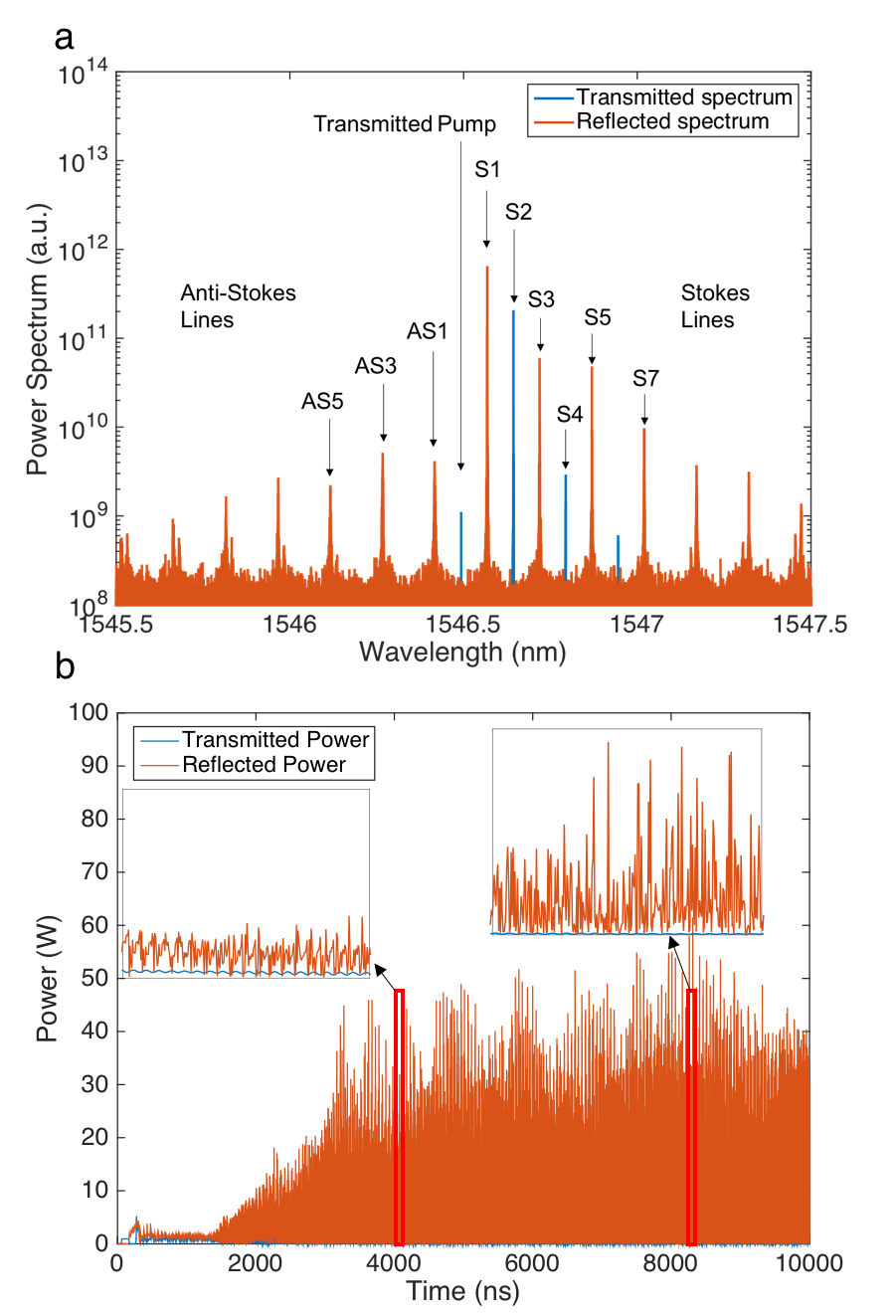}
\end{center}
\caption{(a) Output spectrum of a silica fiber ring cavity with feedback of 99\% of the backward Stokes.  The odd-order Stokes and ant-Stokes lines are spaced by twice the Brillouin frequency.  (b) Temporal output showing erratic pulsations.}
\label{ring_fig}
\end{figure}

Another interesting configuration for cascaded SBS and comb generation is that of a ring-cavity geometry such as the one used by Tang et al to generate a stable frequency comb through SBS and four-wave mixing\cite{Tang2011}.  The experiment used 2.5 km of fiber pumped with 160 mW of power at 1546.5 nm.  The Stokes output was fed back to the input.  Up to 17 lines spaced by twice the Brillouin frequency were observed, with the even orders in the backward direction and the odd orders in the forward direction.  To avoid excessive computation time we shrank the fiber to 10 meters and increased the pump power to 1 W.  We fed back 99\% of the backward first Stokes light to the input.  Figure \ref{ring_fig} shows the result of our simulation using the three equations.  The reflected spectrum contains the odd-order Stokes and anti-Stokes components while the transmitted spectrum contains the depleted pump and the even order Brillouin-shifted components.  In agreement with the experiment, the only strong components in the transmitted spectrum are the pump, second and fourth-order Stokes.  The even transmitted orders higher than fourth are weaker by about 50dB.  The temporal output is erratic, reflecting the absence of a definite phase relationship between the comb lines, a result also noted in the experiment.  The scheme modeled here makes it possible to generate a comb with a spacing of twice the acoustic frequency.  These doubly-spaced frequency components emerged naturally through a mere change of the boundary conditions.  The old approach would have required that we input all the generated Stokes and anti-Stokes waves and their associated acoustic waves and ensured their proper spacing.
 
In order to further test our theory, we consider the experiment of B{\"u}ttner et al in which phase-locking and pulse generation through cascaded SBS and four-wave mixing were observed in a chalcogenide fiber Fabry-Perot cavity\cite{Buttner2014-1}.  In this case the boundary conditions are

\begin{subequations}
\begin{align}
A_+(0,t) = \sqrt{P_{in}(t)}+\sqrt{R}A_-(0,t)\\
A_-(L,t) = \sqrt{R}A_+(L,t)e^{i\Delta\phi_0}
\end{align}
\end{subequations}

\noindent Here the reflectivity $R=22.6\%$, $\Delta \phi_0 = 4\pi n L /\lambda$ is the round trip phase shift at the pump wavelength, the refractive index $n = 2.81$, wavelength  $\lambda = 1550$ nm, and fiber length $L = 38.29$ cm.  The transmitted power is given by 

\begin{align*}
P_{out}(t) = (1-R) |A^+(L,t)|^2
\end{align*}

\begin{figure}[h]
\begin{center}
\includegraphics[scale=1]{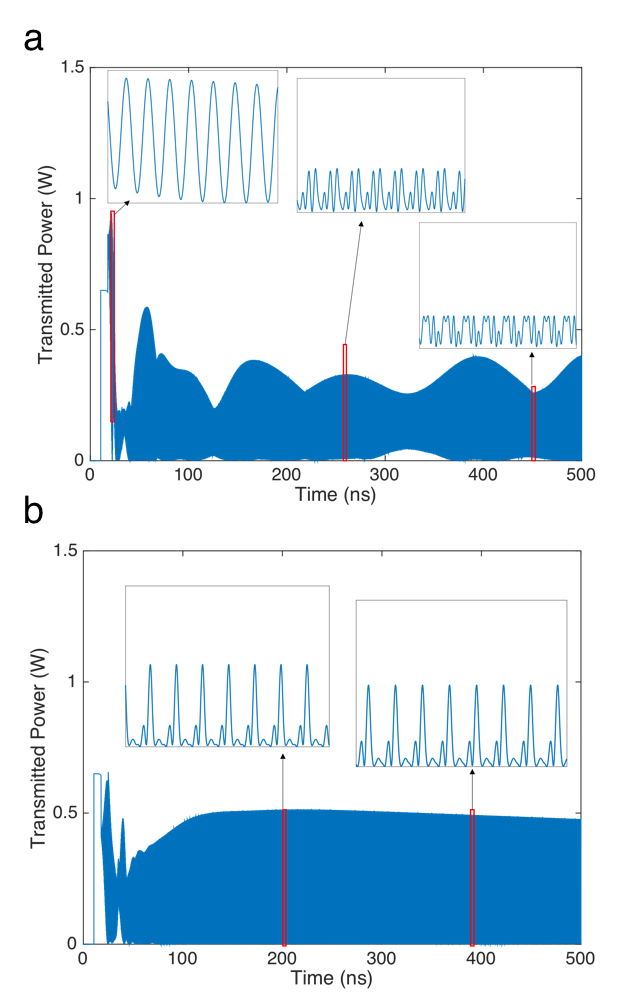}
\end{center}
\caption{Temporal output of 38 cm-long chalcogenide fiber Fabry-Perot cavity for a 500-ns pulse input and two different values of phase shift. a) $\Delta \phi_0 = 1.8\pi$   b) $\Delta \phi_0 = 0.62\pi$}
\label{chalc_fig}
\end{figure}

Figure \ref{chalc_fig} shows the temporal evolution of the output power for two different values of the round trip phase shift $\Delta \phi_0$. In Fig. \ref{chalc_fig}a we see a rapid oscillation with an envelope whose period is about 100 ns, in agreement with the experiment.  The zoomed-in inset taken at around 30 ns shows that the initial rapid oscillations are sinusoidal with a frequency of $\Omega_B/2\pi$ GHz, the Brillouin frequency shift.  They are due to the beating between the pump and the newly-generated first Stokes wave \cite{Buttner2014-1, Ogusu2002-2}.  Later in the evolution of the output pulse, the interference pattern becomes more complicated as higher order Stokes waves are generated.  The oscillation pattern drifts with time indicating that the phase between the interfering waves is not fixed.  In Fig. \ref{chalc_fig}b, for a different value of $\Delta \phi_0$ we find a steady pulsation with a nearly constant envelope.  The nearly constant envelope and the stable pulsation pattern suggest that the modes that interfere to form the pulses are nearly phase locked.  Our simulations are in very good qualitative agreement with the experimental results of Ref. 10.  The power of our approach is that we only needed to solve three equations to capture the dynamics of all the interacting Stokes and anti-Stokes waves.    
\section{Conclusion}

In conclusion, we have introduced a novel approach to studying cascaded stimulated Brillouin scattering and frequency comb generation that requires only three variables and three coupled equations instead of the potentially hundreds that would be needed in the traditional approach.  With this approach the higher-order Stokes and anti-Stokes lines emerge naturally through SBS and four-wave mixing with no need to introduce them as separate waves.  Under the previous treatment of cascaded SBS, finding these lines would require not only many variables and coupled equations, but also knowing \textit{a priori} which frequencies will be amplified with their relative phases. Our new approach automatically picks out the frequencies that are phase-matched, greatly increasing the theory'ss predictive power as we no longer need to manually insert any wave envelopes with predetermined frequencies.  We believe this approach will greatly facilitate the analysis and understanding of cascaded stimulated Brillouin scattering. 

\textit{Acknowledgment}
M. D. thanks the Rackham Graduate School for a 2015 Rackham Summer Award.

\begin{table}[h]
\begin{center}
\begin{tabular}{|c|c|c|}
\hline
Parameter & Chalcogenide Glass & Silica \\ \hline
$g_B$ & $6.1\times10^{-9}$m/W & $1.78\times10^{-11}$m/W \\ \hline
$n_0$  & $2.81$ & $1.50$ \\ \hline
$\Omega_B$ & $2\pi \times 7.8$GHz & $2\pi \times 9.44$GHz \\ \hline
$A_{eff}$ & $56\mu \text{m}^2$ & $11\mu \text{m}^2$ \\ \hline
$\tau_B$ & $12$ns & $3.4$ns \\ \hline
$\gamma$ & $1.737 \text{(Wm)}^{-1}$ & $11.5\times10^{-3} \text{(Wm)}^{-1}$  \\ \hline
$\alpha$ & $0.1935 \text{m}^{-1}$ & $0.0064 \text{m}^{-1}$  \\ \hline
\end{tabular}
\end{center}
\caption{Parameter values used in the simulations}
\label{table_param}
\end{table}

\bibliography{prl_draft}

\end{document}